\documentclass[preprint,prd,showpacs,floatfix]{revtex4}

\usepackage{amsmath}
\usepackage{amsfonts}
\usepackage{amssymb}
\usepackage{bbm}
\usepackage{latexsym}
\usepackage[dvips]{graphicx}
\usepackage{color}

\begin{document}

\title{Towards testing interacting cosmology by distant type Ia supernovae} 
\author{Marek Szyd{\l}owski}
\email{uoszydlo@cyf-kr.edu.pl}
\affiliation{Astronomical Observatory, Jagiellonian University, 
Orla 171, 30-244 Krak\'ow, Poland}
\author{Tomasz Stachowiak}
\affiliation{Astronomical Observatory, Jagiellonian University, 
Orla 171, 30-244 Krak\'ow, Poland}
\author{Rados{\l}aw Wojtak}
\affiliation{Astronomical Observatory, Jagiellonian University, 
Orla 171, 30-244 Krak\'ow, Poland}
\date{\today}

\begin{abstract}
We investigate the possibility of testing cosmological models with interaction 
between matter and energy sector. We assume the standard FRW model while the 
so called energy conservation condition is interpreted locally in terms 
of energy transfer. We analyze two forms of dark energy sectors: the 
cosmological constant and phantom field. We find a simple exact solution of 
the models in which energy transfer is described by a Cardassian like term in 
the relation of $H^{2}(z)$, where $H$ is Hubble's function and $z$ is redshift. 
The considered models have two additional parameters $(\Omega_{\text{int}},n)$ 
(apart the parameters of the $\Lambda$CDM model) which can be tested using 
SNIa data. In the estimation of the model parameters Riess et al.'s sample is 
used. We also confront the quality of statistical fits for both the 
$\Lambda$CDM model and the interacting models with the help of the Akaike 
and Bayesian informative criteria. Our conclusion from standard best fit
method is that the interacting models explains the acceleration of the 
Universe better but they give rise to a universe with high matter density. 
However, using the tools of information criteria we find that the two new 
parameters play an insufficient role in improving the fit to SNIa data and the 
standard $\Lambda$CDM model is still preferred. We conclude that high precision 
detection of high redshift supernovae could supply data capable of justifying
adoption of new parameters. 
\end{abstract}

\maketitle

\section{Introduction}

The main aim of the paper is to use the distant type Ia supernovae (SNIa) 
compiled by Riess et al. \cite{Riess:2004nr} to analyze the interacting 
cosmology (or cosmology with energy transfer) which was recently proposed by 
Zimdahl and Pavon \cite{Zimdahl:2003wg} and Zhang \cite{Zhang:2005rj} as a 
way to explain the coincidence conundrum of the ratio of dark matter to dark 
energy density at present. We show that allowed intervals for two independent 
parameters which characterize interacting cosmology give rise to high matter 
density, which can hardly be reconciled with the current value obtained from 
cluster baryon fraction measurements. 

The major development of modern cosmology, which mainly concentrates on 
observational cosmology, seems to be a discovery of the acceleration of 
the universe through distant SNIa. In principal there are three different 
theoretical types of explanation of acceleration. In the first group the 
existence of matter of unknown forms violating the strong energy condition is 
postulated (dark energy). In the second group instead of dark energy 
some modification of the FRW equation is proposed. Between these two 
propositions lies an idea of interacting cosmology in which the source 
of acceleration lies in interaction between dark energy and dark matter 
sector. 

The starting point of any investigations in observational cosmology is the 
naive FRW model (the $\Lambda$CDM model) in which matter and dark energy 
sectors do not interact with each other. The next step in making this model 
more realistic is to include the interaction between dark matter and dark 
energy. 
We start with the FRW standard cosmological equations: the Raychaundhuri 
equation and conservation condition. They assume the following forms 
respectively 
\begin{align}
\frac{\ddot{a}}{a} &= - \frac{1}{6} (\rho + 3p) \label{eq:1} \\
\dot{\rho} &= -3H( \rho+p ), \label{eq:2}
\end{align}
where $a$ is the scale factor, $\rho$ and $p$ are energy density and pressure 
of perfect fluid, a source of gravity; $H=\dot{a}/a$ is the Hubble function, 
and a dot means differentiation with respect to the cosmological time $t$. It 
would be convenient to rewrite relation (\ref{eq:2}) for mixture of dust 
matter and dark energy
\begin{equation}
\label{eq:3}
\frac{1}{a^{3}}\frac{d}{dt} \left(\rho_{\text{m}}a^{3}\right) +
\frac{1}{a^{3(1+w_{X})}}\frac{d}{dt}\left( \rho_{X}a^{3(1+w_{X})}\right)=0,
\end{equation}
where the total pressure and energy density are
\begin{align}
\label{eq:4}
p &= 0+w_{X}\rho_{X},\quad w_{X}=p_{X}/\rho_{X}=\text{const} \\
\label{eq:5}
\rho &= \rho_{\text{m}}+\rho_{X}
\end{align}
The standard postulate of the $\Lambda$CDM model is that both terms in 
eq.~(\ref{eq:3}) are equal to zero, or that the matter components evolve
independently. We, however, assume the existence of interaction introducing 
a non-vanishing function $\gamma(t)$ such that eq.~(\ref{eq:3}) is still 
satisfied locally
\begin{align}
\label{eq:6}
\frac{1}{a^{3}}\frac{d}{dt}\left(\rho_{\text{m}}a^{3}\right) &\equiv \gamma(t) \\
\label{eq:7}
\frac{1}{a^{3(1+w_{X})}}\frac{d}{dt}\left( \rho_{X}a^{3(1+w_{X})} \right) 
&\equiv -\gamma(t)
\end{align}
Note that $\gamma(t)$ in (\ref{eq:6})-(\ref{eq:7}) has the interpretation 
of the rate of change of energy in the unit comoving volume. 

We only assume the existence of the energy transfer without specifying its 
physical mechanism, then test it by astronomical observations. The idea of 
such decomposition of the conservation condition (\ref{eq:6})-(\ref{eq:7}) is 
not new \cite{Zimdahl:2001ar}. The interacting quintessence models with 
pressureless cold dark matter fluid can give rise to the scale factor relation 
$\rho_{X}/\rho_{\text{m}}=a^{\xi}$ \cite{Hoffman:2003ru,Pavon:2004xk}, where 
$\xi$ is a constant parameter. This relation was also tested by CMB WMAP 
observations and statistically investigated in the context of SNIa 
observations \cite{Zimdahl:2002zb} but with the help of a different exact 
solution. In the paper of one of us \cite{Szydlowski:2005ph}, this idea was 
investigated independently without any physical assumption or about the form 
of such an interaction or an exact solution. Under such general assumptions, 
that the FRW dynamics is valid, the existence of energy transfer to the dark 
energy sector from the dark matter section is favored by SNIa data. However, 
if we apply the prior $\Omega_{\text{m},0} = 0.3$ as claimed by independent 
galactic observations, then the transfer between dark matter and phantom 
sector is required. The proposition of different physical mechanism of energy 
exchange from dark energy to dark matter sectors was proposed by many authors 
\cite{Ziaeepour:2003qs,Svetlichny:2005ke,Zhang:2005jj}. 

We consider models where the rate of energy transfer is proportional to the 
rate of change of the scale factor $\gamma(t)=f(a)H$.

While the standard statistical methods of maximum likelihood used for 
estimation of the model parameters favor the energy transfer between 
dark matter and phantom dark energy sector, we demonstrate by the means of 
the Akaike and Bayesian information criteria that the $\Lambda$CDM 
model with interaction is selected.

\section{The FRW model with interaction}

We assume for simplicity of presentation without loss of generality a simple 
flat model for which we have the Friedmann first integral in the form
\begin{equation}
\label{eq:8}
3\dot{a}^{2}=a^{2}(\rho_{\text{m}}+\rho_{X})
\end{equation}
In the case of non-flat models relation (\ref{eq:8}) is of the form
\begin{equation}
\label{eq:9}
3\dot{a}^{2}=a^{2}(\rho_{\text{m}}+\rho_{X})-3k
\end{equation}
We add to equation (\ref{eq:8}) the generalized adiabatic condition
\begin{align}
\frac{d}{dt}\rho_{\text{m}} &= -3H\rho_{\text{m}}+\gamma(t), \label{eq:10} \\
\frac{d}{dt}\rho_{X} &= -3(1+w_{X})H\rho_{X}-\gamma(t), \label{eq:11} \\
\frac{d}{dt}H &= - H^2 - \frac{1}{6} (\rho_{\text{m}} + \rho_{X}) \label{eq:12}
\end{align}
from which we obtain
\begin{align}
\frac{d(\ln \rho_{\text{m}})}{d(\ln a)} &\equiv I_{a}(\rho_{\text{m}}) 
= -3+\frac{\gamma}{H\rho_{\text{m}}} \label{eq:13} \\
\frac{d(\ln \rho_{x})}{d(\ln a)} &\equiv I_{a}(\rho_{X})
= -3(1+w_{X})-\frac{\gamma}{H\rho_{X}} \label{eq:14} \\
\frac{d(\ln(\rho_{\text{m}}/\rho_{X})}{d(\ln a)} &= 
3w_{X} + \frac{1}{H} \left( \frac{1}{\rho_{\text{m}}} + \frac{1}{\rho_{X}} 
\right) \label{eq:15}
\end{align}
where $I_{a}(\rho_{\text{m}})$ and $I_{a}(\rho_{X})$ have the interpretation 
of elasticity (logarithmic slope) of the energy densities with respect to the 
scale factor. If energy transfer vanishes then the corresponding elasticity 
function becomes constant. From (\ref{eq:13})-(\ref{eq:14}) we can simply 
derive the relation 
\begin{equation}
\label{eq:16}
I_{a}\left( \frac{\rho_{\text{m}}}{\rho_{X}} \right) 
= \frac{d\ln \left(\frac{\rho_{\text{m}}}{\rho_{X}}\right)}
{d(\ln a)}=3w_{X}+\frac{3\gamma H}{\rho_{\text{m}}\rho_{X}}
\end{equation}
Therefore scaling solutions $\rho_{\text{m}}/\rho_{X}\propto a^{\xi}$ can 
only be realized if a very special form of parameterization $\gamma(t)$ is 
assumed \cite{Zimdahl:2001ar,Pavon:2004xk,Zimdahl:2002zb}, namely
\begin{equation}
\label{eq:17}
\gamma(t)=\gamma_{0} H \frac{1}{\frac{1}{\rho_{\text{m}}}+\frac{1}{\rho_{X}}}
=\gamma_{0}\bar{\rho}\,^{-1}H,
\end{equation}
where $\bar{\rho}$ is the harmonic average of $\rho_{\text{m}}$ and $\rho_{X}$. 
It is interesting to discuss the existence of scaling solutions for other 
parameterizations of $\gamma(t)$. For this purpose the methods of dynamical 
systems can be useful because they offer possibility of investigating all 
admissible evolutional paths for all initial conditions. Let us consider the 
following distinguished case.

Let $\gamma(t)=\gamma_{0}f(a)H$ and $f(a)=a^{m}$. In this case one can find 
exact solutions determining energy densities 
\begin{align}
\label{eq:18}
\rho_{\text{m}} &= \frac{\gamma_{0}}{m+5}a^{m+2}+\frac{A_{\text{m},0}}{a^{3}} \\
\label{eq:19}
\rho_{X} &= -\frac{\gamma_{0}}{m+5+3w_{X}}a^{m+2}
+\frac{A_{\text{m},0}}{a^{3(1+w_{X})}}
\end{align}
where in the special cases of $m=-5$ and $m=-5-3w_{X}$ instead of a power law 
function a logarithmic function appears; $A_{\text{m},0}$ and $A_{X,0}$ are 
constants. Note that the contribution to the total energy arising from 
interaction is of the type $a^{-m-2}$, therefore it can be modelled by some 
additional fictitious fluid satisfying the equation of state 
$p_{\text{int}} = -(m+5)\rho_{\text{int}}/3$. 

In terms of redshift $z$ we obtain the basic relation
\begin{equation}
\label{eq:20}
H^{2}=H_{0}^{2}\left( \Omega_{\text{int}}(1+z)^{-m-2}
+\Omega_{\text{m},0}(1+z)^{3}+\Omega_{X,0}(1+z)^{3(1+w_{X})}
+\Omega_{k,0}(1+z)^{2} \right),
\end{equation}
where
\begin{equation}
\Omega_{int,0}=\frac{3w_{X}}{(m+5)(m+5+3w_{X})}\gamma_0,
\end{equation}
and the index zero means that quantities are evaluated at the present 
moment. The relation will be useful in the next section where both 
$\Omega_{\text{int}}$ and $m$ parameters will be fitted to SNIa data. 

\begin{figure}
\includegraphics[width=0.65\textwidth]{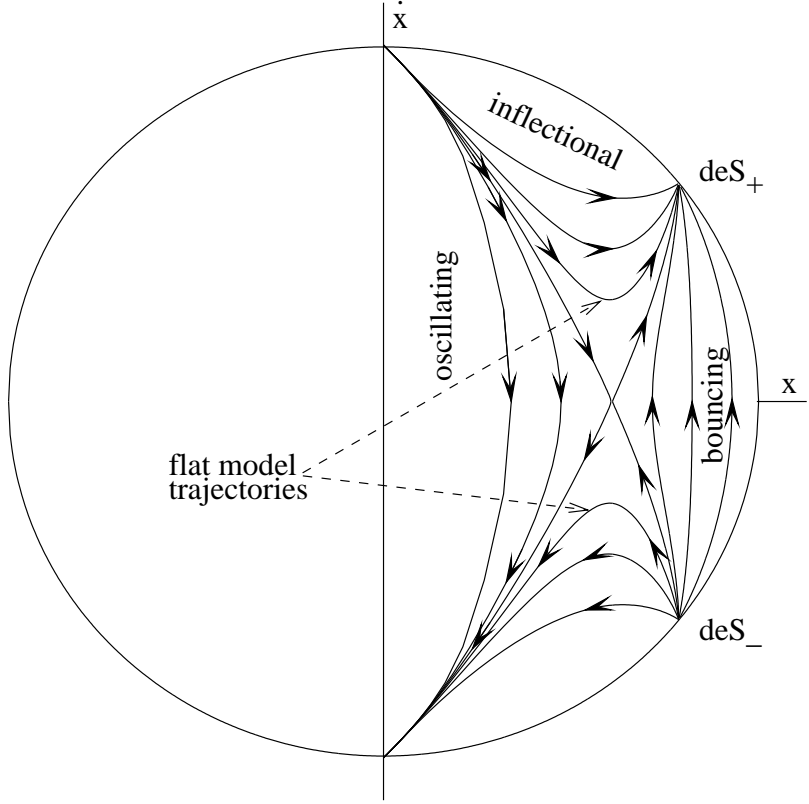}
\caption{The phase portrait for the $\Lambda$ interacting cosmological model.
The trajectories of flat model separate the domains of closed and open models}
\label{fig:1}
\end{figure}

\begin{figure}
\includegraphics[width=0.65\textwidth]{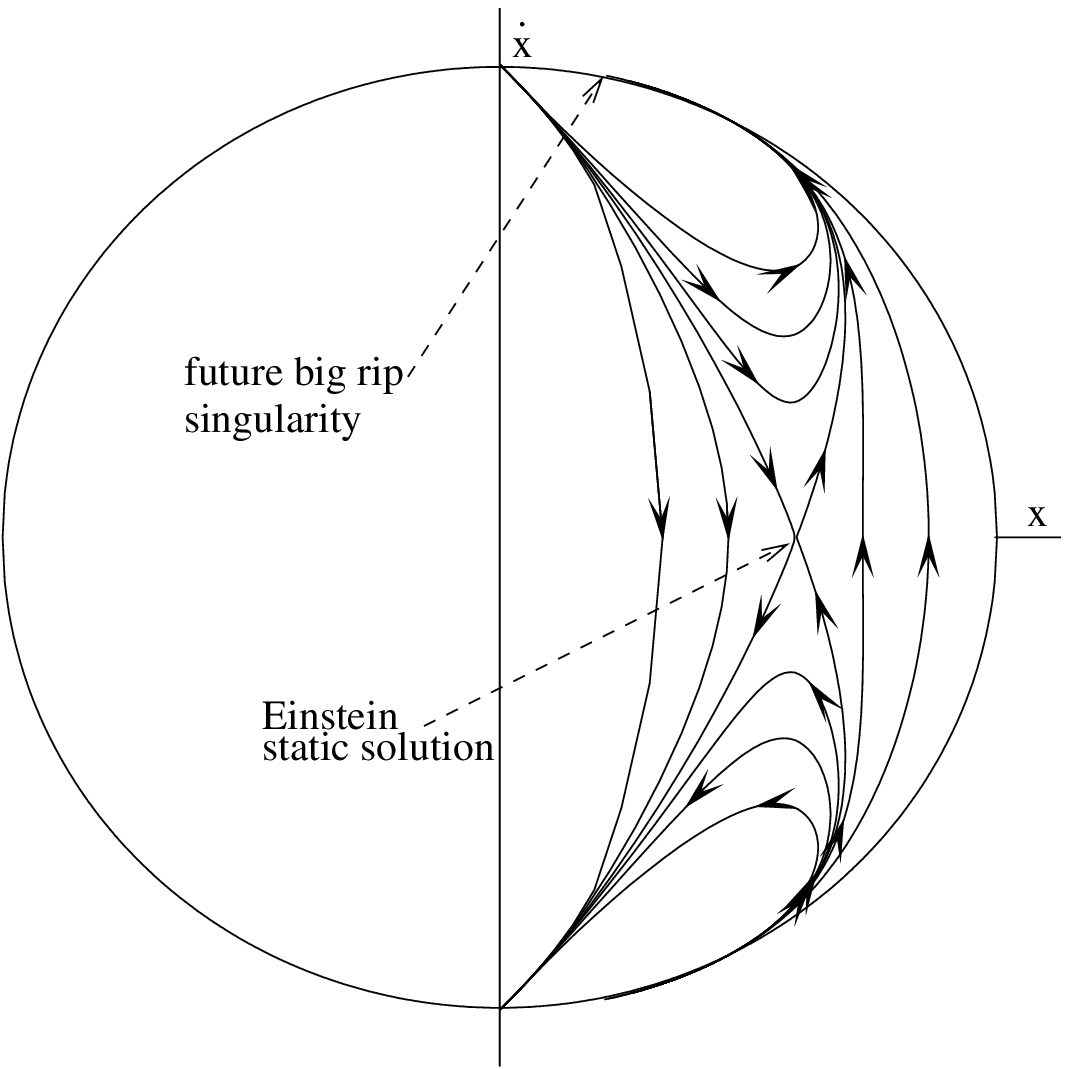}
\caption{The phase portrait for the phantom interacting cosmological model.}
\label{fig:2}
\end{figure}

For formulation the corresponding dynamical system it would be useful to 
represent this case in the form of the Hamiltonian dynamical system. For this 
case it is sufficient to construct the potential function
\begin{equation}
\label{eq:21}
V=-\frac{\rho_{\text{eff}}a^{2}}{6}
\end{equation}
which build the Hamiltonian
\begin{equation}
\label{eq:22}
\mathcal{H}=\frac{\dot{a}^{2}}{2}+V(a).
\end{equation}
We find that the Hamiltonian system is determined on the zero level 
$\mathcal{H}=0$ and the potential function in terms of the scale factor 
expressed in units of its present value $a_0$, $x=\frac{a}{a_0}$, is
\begin{equation}
\label{eq:23}
V(x)=-\frac{1}{2}\Omega_{\text{m},0}x^{-1}
-\frac{1}{2}\Omega_{w_{X},0}x^{-3(1+w_{X})+2}
-\frac{1}{2}\Omega_{\text{int}}x^{-m}-\frac{1}{2}\Omega_{k,0}.
\end{equation}
The Hamiltonian constraint gives rise to the relation 
$\sum_{i}\Omega_{i,0}=1$. The corresponding dynamical system is of the form
\begin{align}
\dot{x} &= y \label{eq:24} \\
\dot{y} &= -\frac{\partial V}{\partial x} \label{eq:25}
\end{align}
and $\mathcal{H} = 0$ plays the role of first integral of 
(\ref{eq:24})-(\ref{eq:25}). Advantages of representing dynamics in the form 
(\ref{eq:24})-(\ref{eq:25}) is that the critical points as well as their 
character can be determined from the geometry of the potential function only. 
Only the static critical points, admissible in the finite domain, 
$y_{0}=0|_{x_0}$, $\frac{\partial V}{\partial x}=0$ are saddles or centers. 
The eigenvalues of the linearization matrix are 
\begin{equation}
\label{eq:26}
\lambda_{1,2} = \pm \sqrt{- \left( \frac{\partial^{2}V}{\partial x^{2}} 
\right) _{x_{0}}}
\end{equation}
The phase portraits on the phase plane $(x,\dot{x})$ are shown in Fig. 
\ref{fig:1} ($w=-1$, ``decaying'' vacuum cosmology) and in Fig.~\ref{fig:2} 
($w=-\frac{4}{3}$, phantom fields). We can observe the existence of a single 
static critical point located on the $x$-axis and the topological equivalence 
of these phase portraits in the finite domain. In both cases we have static 
critical points situated on the $x$-axis. They are of saddle type because the 
potential function is upper convex. We have also the critical points at 
infinity which represent stable nodes. They represent the de Sitter model 
in Fig.~\ref{fig:1} and the big rip singularity in Fig.~\ref{fig:2}. The big 
rip singularity is generic future of the model violating weak energy condition 
at which the scale factor is infinite at some finite time 
\cite{Dabrowski:2003jm}.

In connection with the present value of the density parameters 
$\Omega_{\text{m},0}$ and $\Omega_{X,0}$ there arises the coincidence 
problem. Why are the energy densities of dark energy and dark matter 
of the same order amplitude at the present epoch? 
\cite{Dalal:2001dt,Sahni:2002kh,Gorini:2004by,Huey:2004qv}. In this context 
we can use some interesting idea formulated by Amendola and Tocchini-Valentini 
\cite{Amendola:2000uh} who argue that the present state of the accelerating 
universe is described by the global attractor in the phase space. It seems to 
be reasonable to treat the evolution of the interacting universe model 
with tools of dynamical system methods. From the physical point of view 
the critical points represent asymptotic state of the system reached 
in infinite time. Let us assume that $\gamma(t)$ is proportional to 
the Hubble parameter, i.e. $\gamma(t)\propto \bar{\gamma}(\rho_{\text{m}}, 
\rho_{X})H$ \cite{Zimdahl:2001ar,Pavon:2004xk,Zimdahl:2002zb} then 
after the reparametrization of the time variable $t \to \tau \colon 
Hdt = d\tau$, we obtain the two-dimensional closed dynamical system 
in the form 
\begin{align}
\label{eq:27}
\frac{d\rho_{\text{m}}}{d\tau} &= - 3\rho_{\text{m}} + 
\bar{\gamma}(\rho_{\text{m}}, \rho_{X}) \\
\label{eq:28}
\frac{d\rho_{X}}{d\tau} &= - 3(1+w) - 
\bar{\gamma}(\rho_{\text{m}}, \rho_{X}). 
\end{align}
System (\ref{eq:27})-(\ref{eq:28}) has a critical point at which 
\[
\Omega_{X,0} / \Omega_{\text{m},0} = - (1+w)^{-1}.
\]
In the case of phantom ($w=-4/3$) we obtain 
$\Omega_{X,0} / \Omega_{\text{m},0} = 3$. This critical point is stable 
if a trace of a linearization matrix is negative at this point, i.e., 
$\partial(-\bar{\gamma})/\partial\rho_{\text{m}} > 
\partial(-\bar{\gamma})/\partial\rho_{X} - 2$. 
If $\bar{\gamma} \propto - \rho_{\text{m}}\rho_{X}$ it means that 
$\Omega_{X,0} > \Omega_{\text{m},0}$.

\begin{figure}
\includegraphics[width=0.65\textwidth]{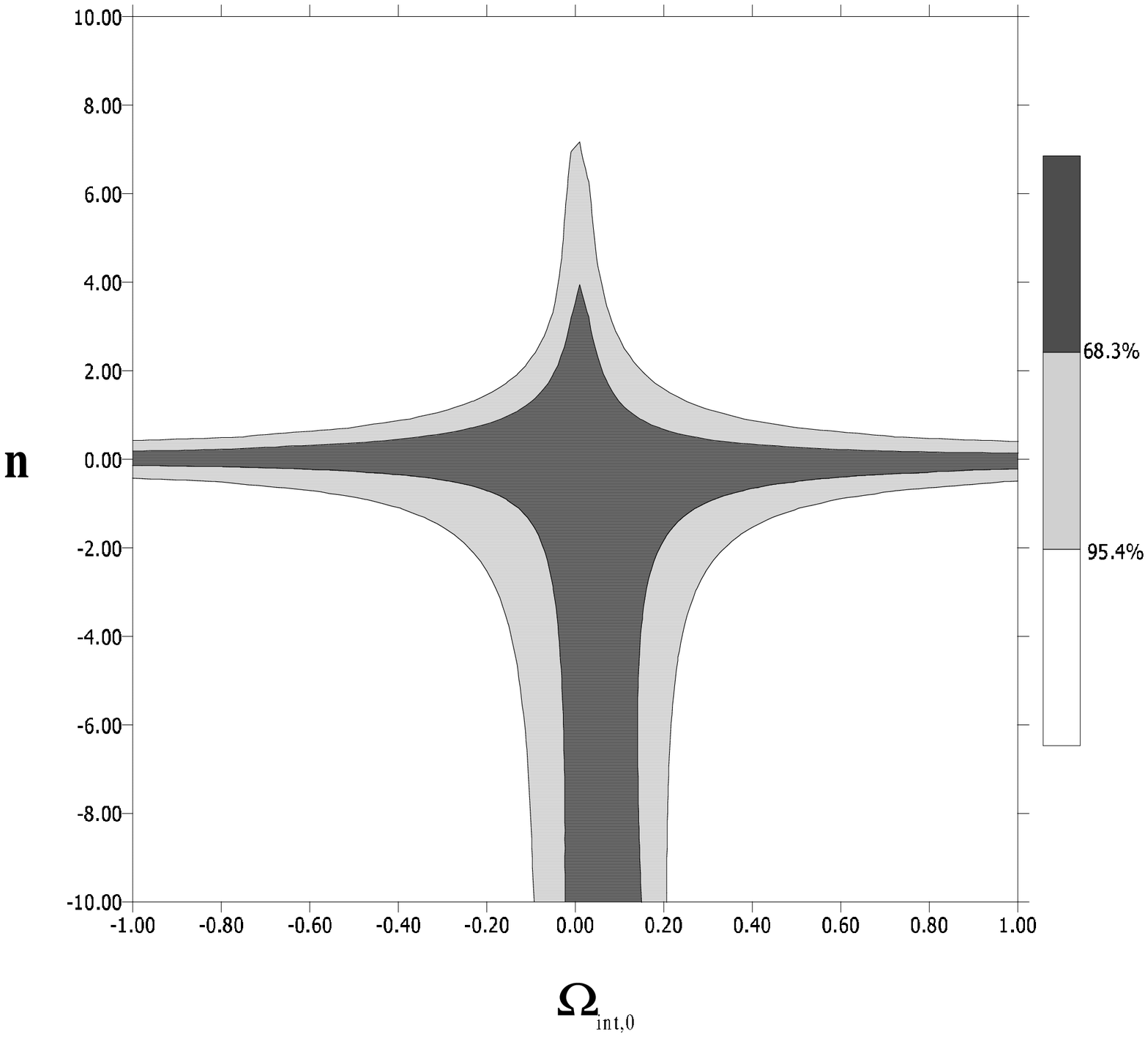}
\caption{Confidence levels on the plane 
$(\Omega_{\text{int}},n)$ for the interaction between matter and 
the $\Lambda$ vacuum energy sectors; the prior on $\Omega_{\text{m},0}=0.3$ 
is assumed.}
\label{fig:3}
\end{figure}

\begin{figure}
\includegraphics[width=0.65\textwidth]{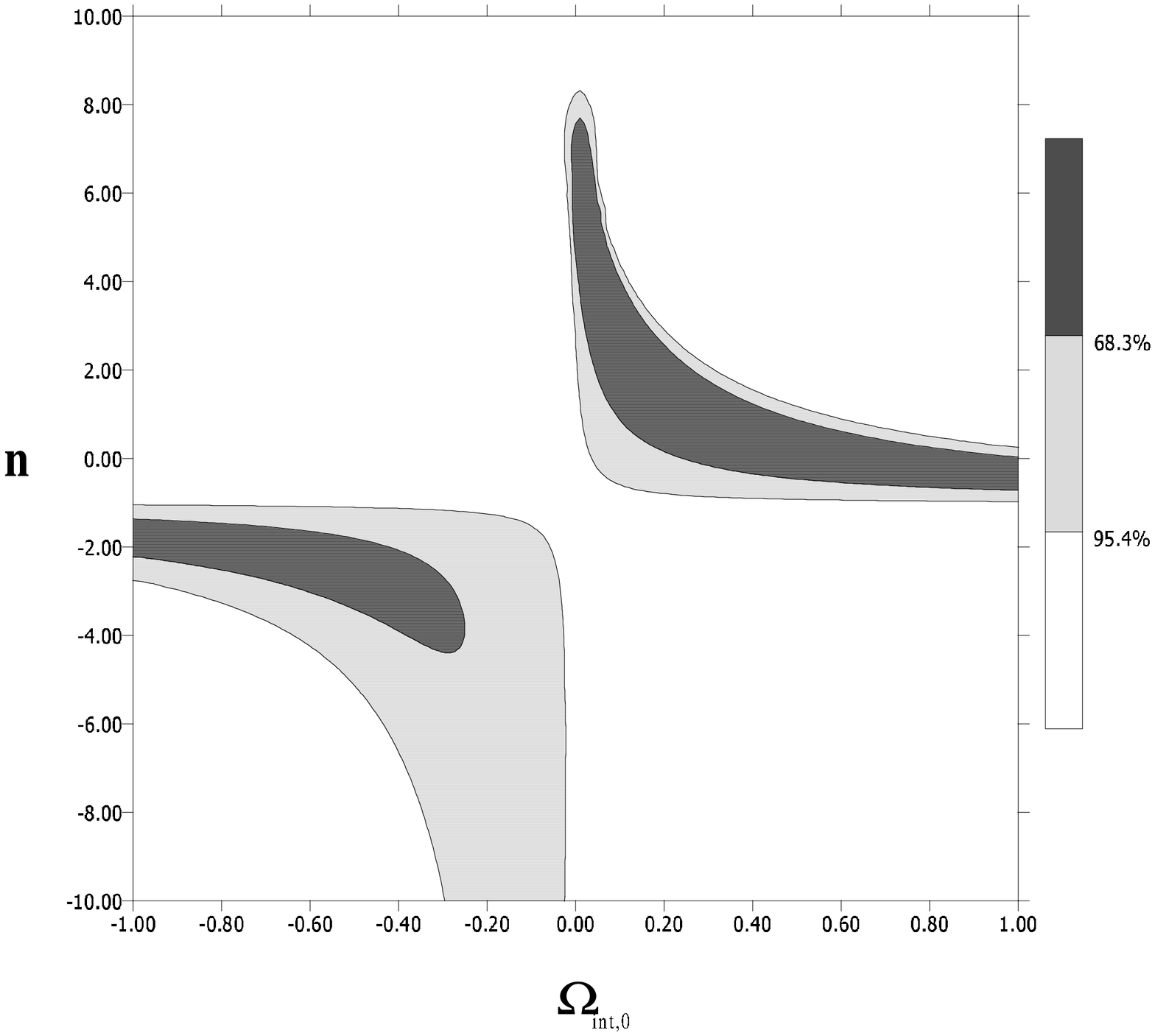}
\caption{Confidence levels on the plane 
$(\Omega_{\text{int}},n)$ for the interaction between matter and 
the phantom sectors; the prior on $\Omega_{\text{m},0}=0.3$ is assumed.}
\label{fig:4}
\end{figure}

In Fig.~\ref{fig:3} we observe that the most probable values of $n=-m-2$ and 
$\Omega_{\text{int},0}$ are close to zero. In Fig.~\ref{fig:4} there is two 
disjoint region which are preferred, while values of $n$ and 
$\Omega_{\text{int},0}$ close to zero are excluded at the 95\% confidence 
level.

\begin{figure}
\includegraphics[width=0.4\textwidth]{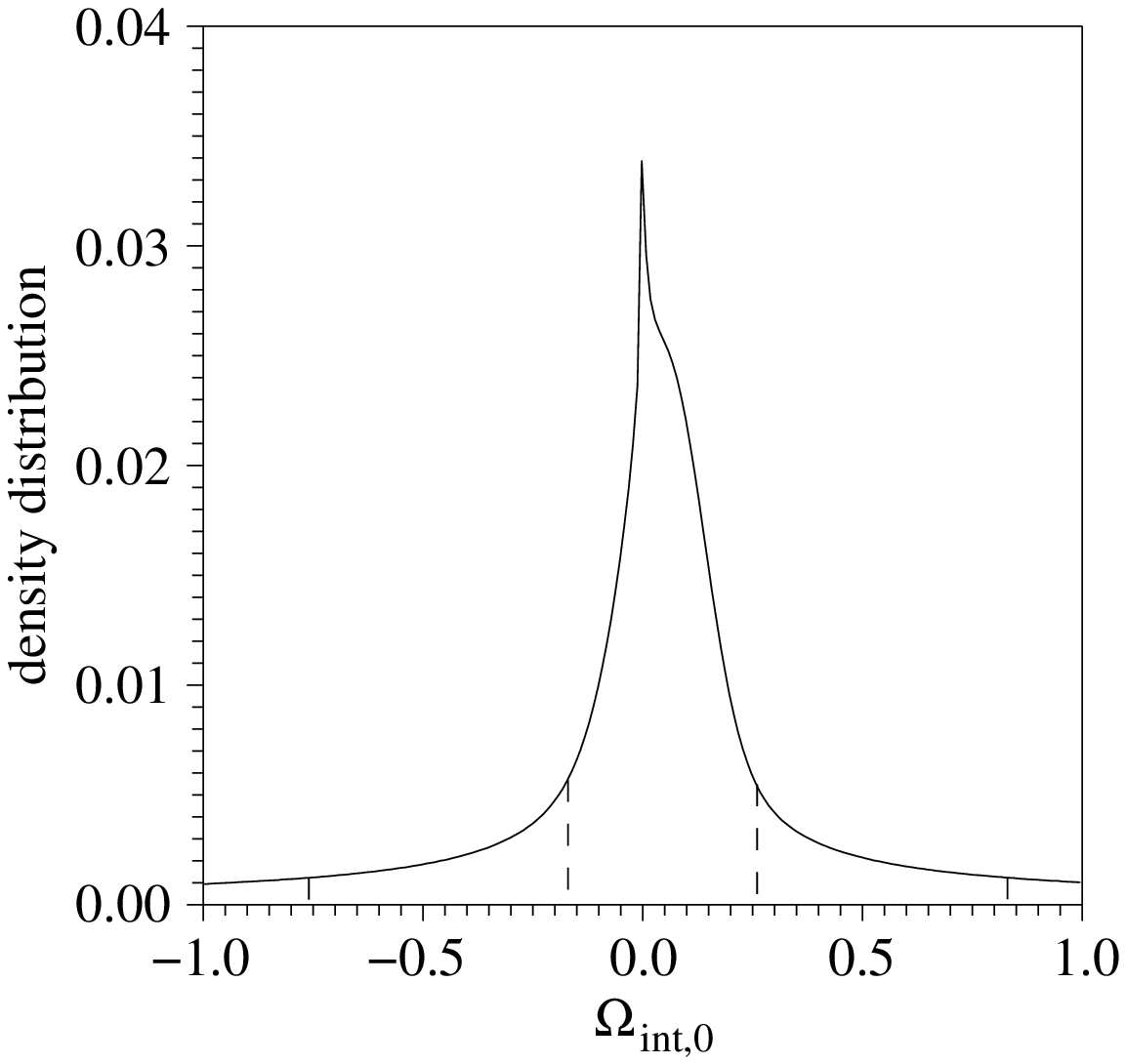}
\includegraphics[width=0.4\textwidth]{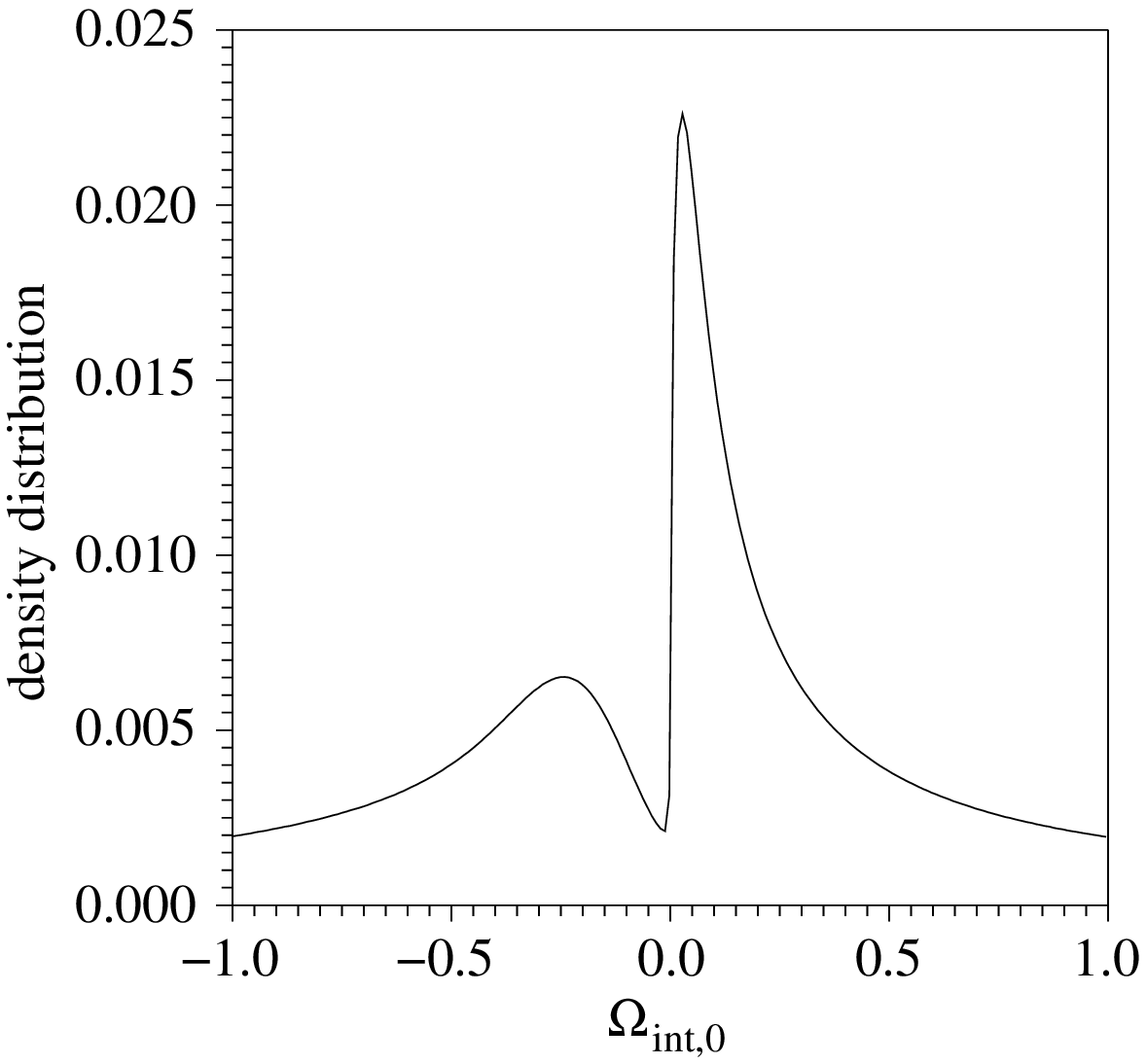}
\caption{
The PDF for $(\Omega_{\text{int},0})$ when matter interacts with the 
``decaying'' vacuum (left) and  matter interacts with phantom field (right);
the prior on $\Omega_{\text{m},0}=0.3$ is assumed.}
\label{fig:5}
\end{figure}

\begin{figure}
\includegraphics[width=0.4\textwidth]{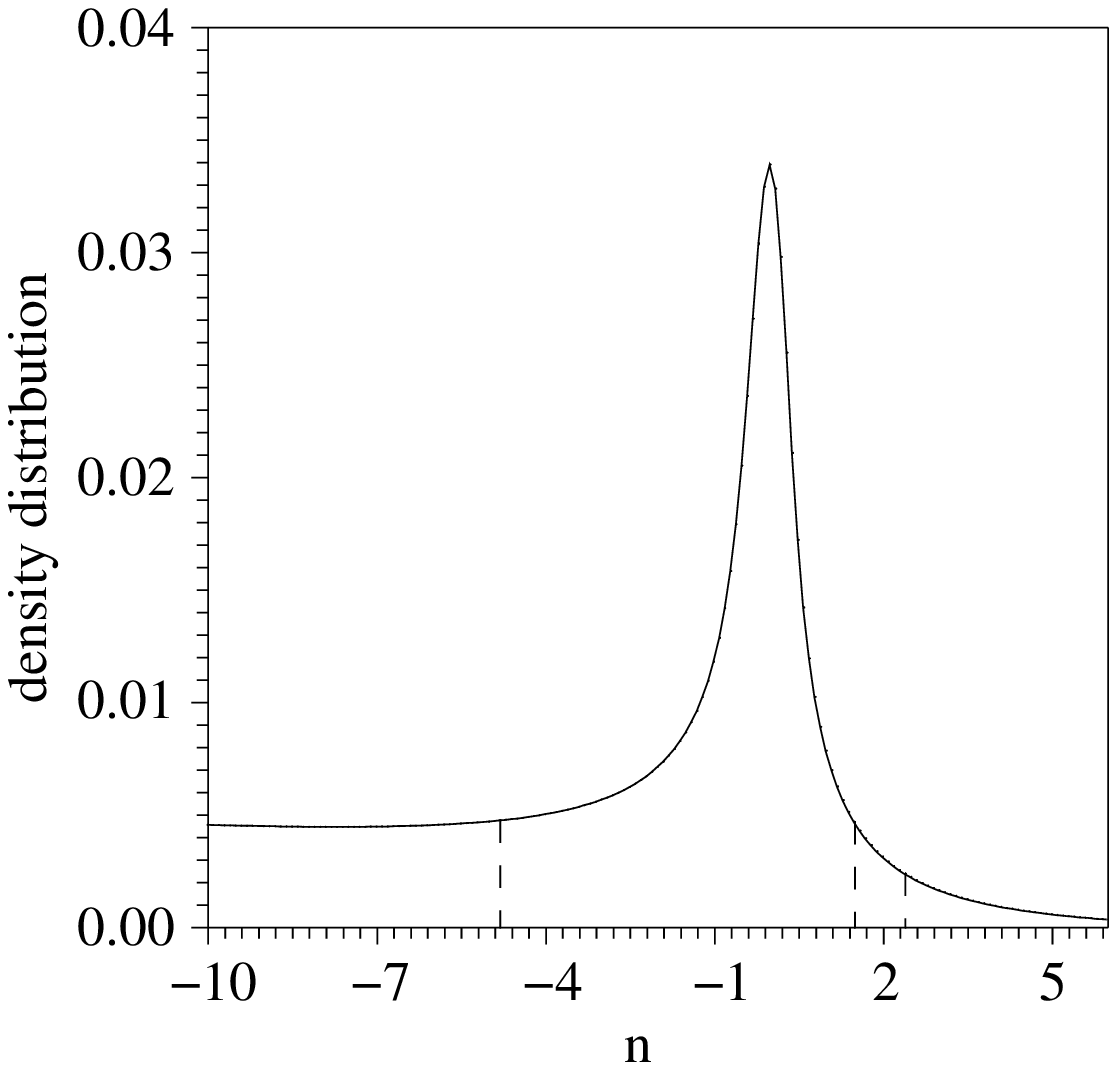}
\includegraphics[width=0.4\textwidth]{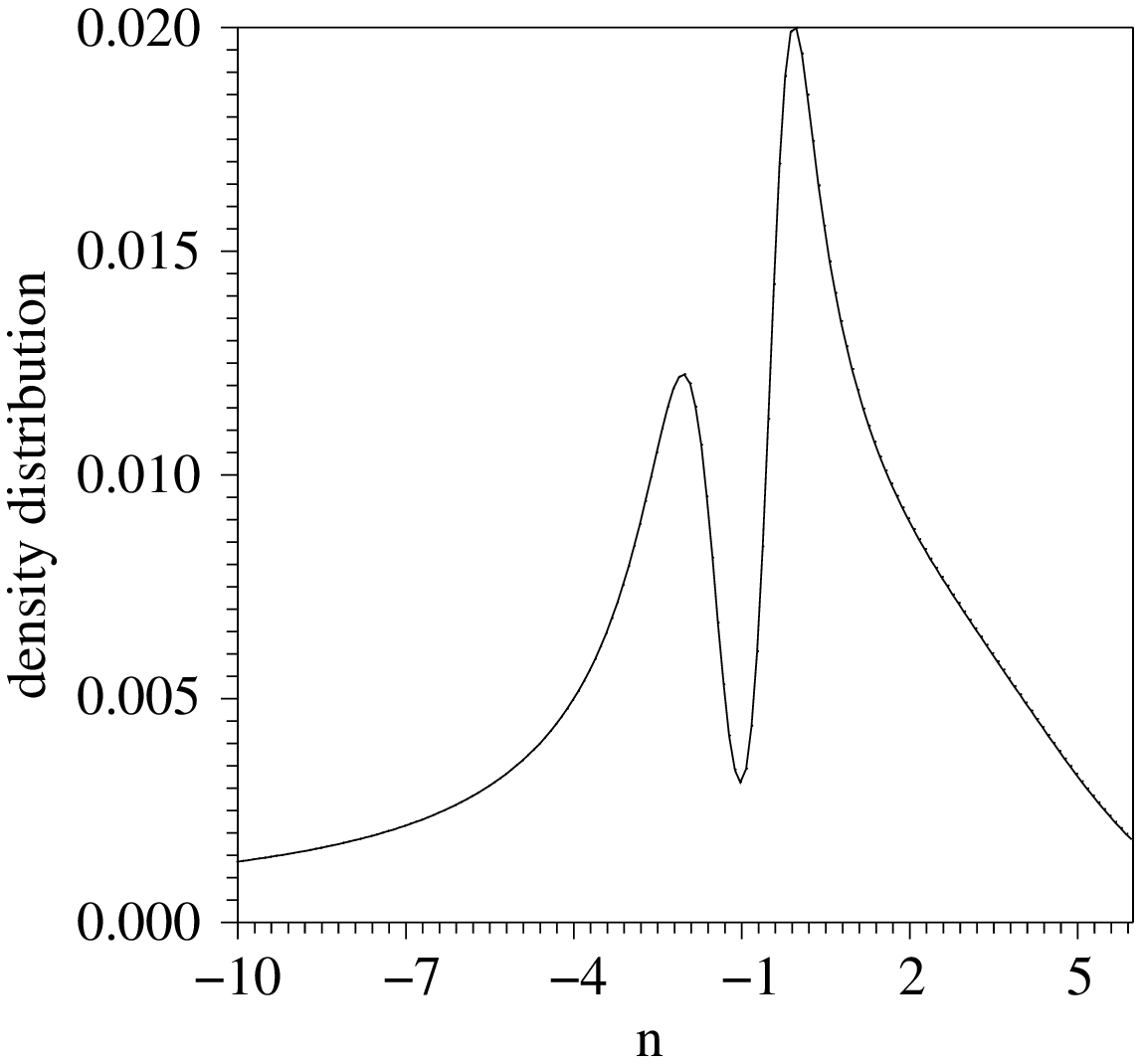}
\caption{
The PDF for $n$ when matter interacts with the 
``decaying'' vacuum (left) and  matter interacts with phantom field (right);
the prior on $\Omega_{\text{m},0}=0.3$ is assumed.}
\label{fig:6}
\end{figure}

\begin{figure}
\includegraphics[width=0.4\textwidth]{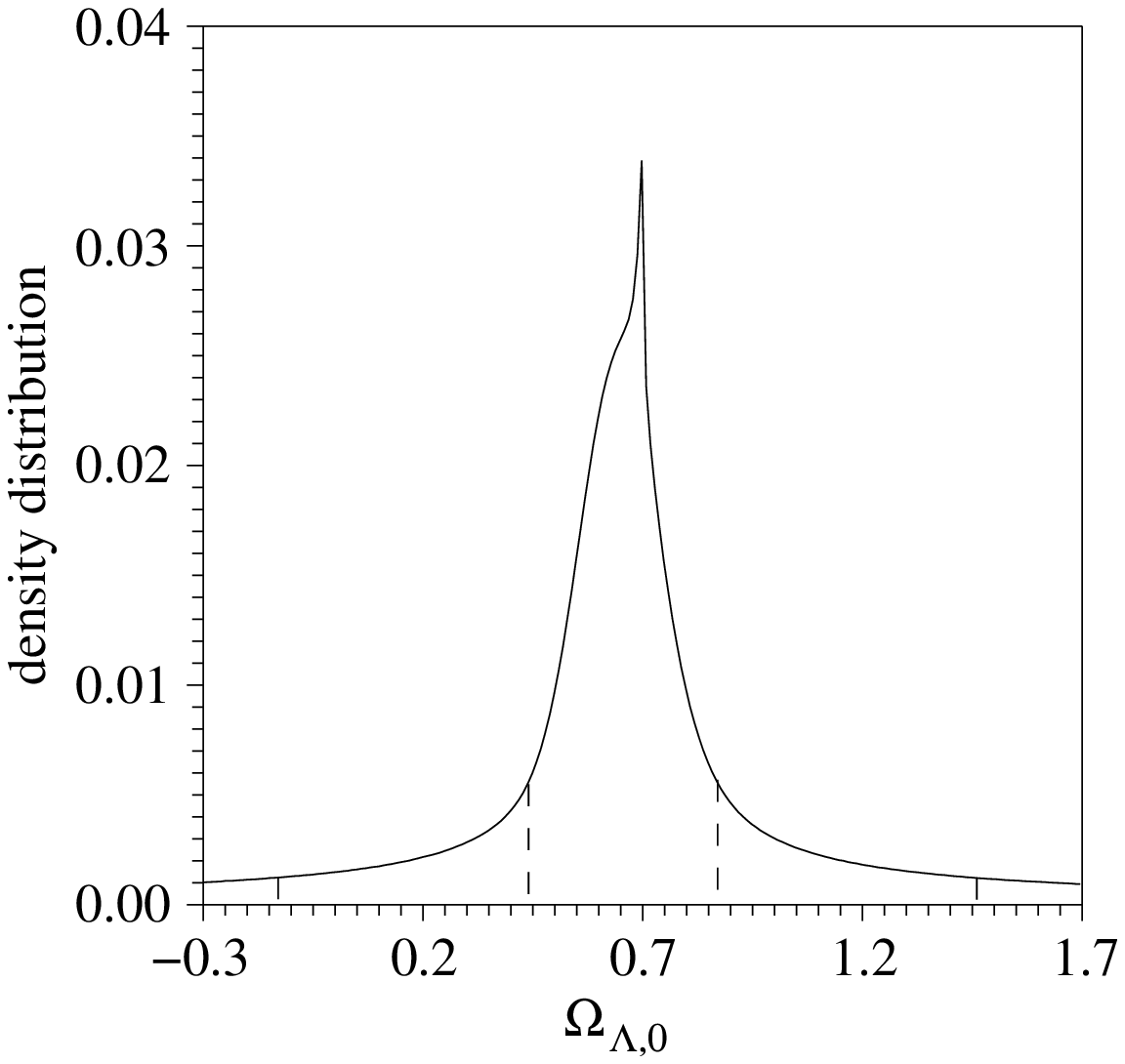}
\includegraphics[width=0.4\textwidth]{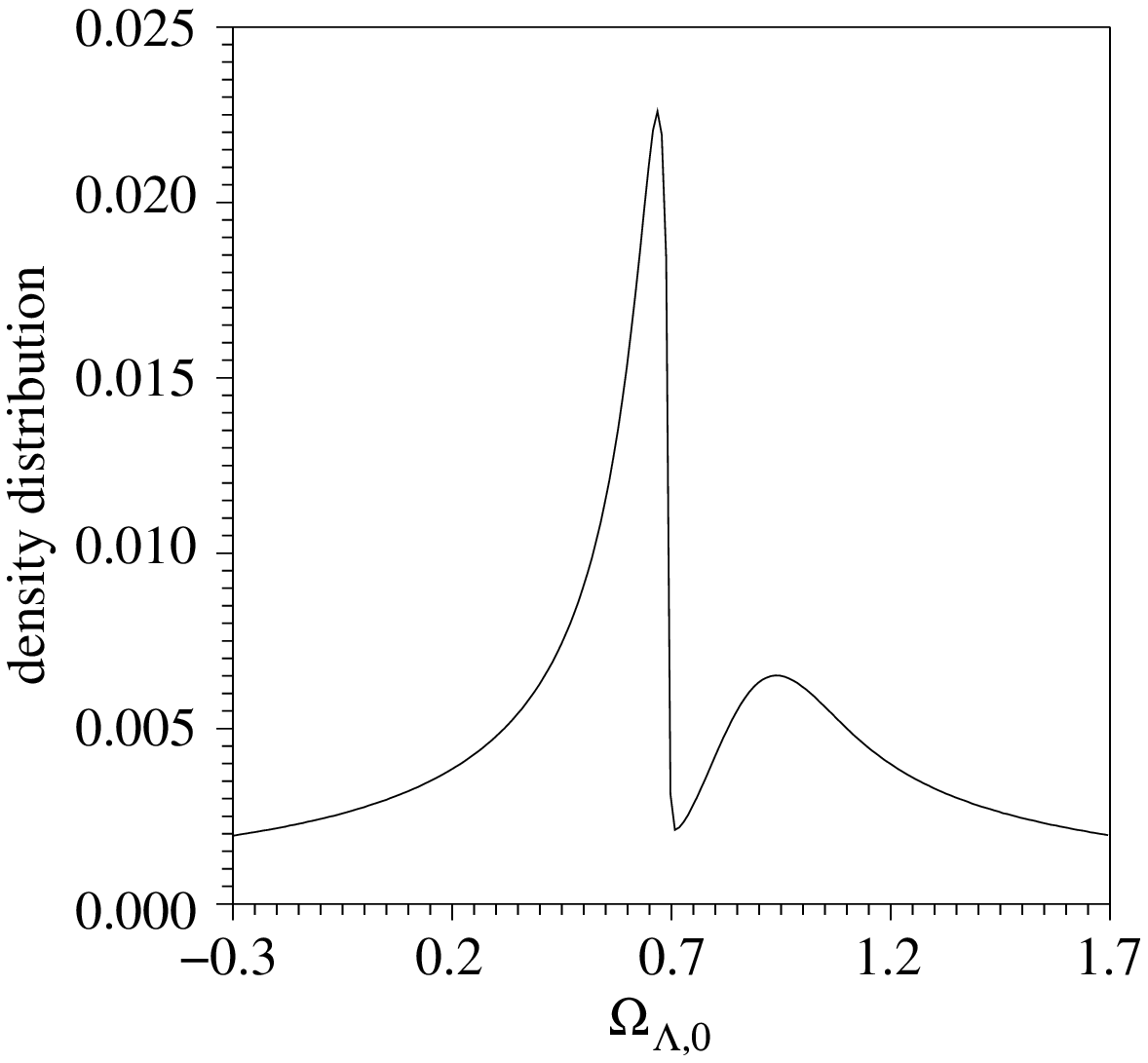}
\caption{
The PDF for $(\Omega_{\Lambda,0})$ when matter interacts with the 
``decaying'' vacuum (left) and  matter interacts with phantom field (right); 
the prior on $\Omega_{\text{m},0}=0.3$ is assumed}
\label{fig:7}
\end{figure}

For deeper statistical analysis it is useful to consider the probability 
distribution function (PDF) over model parameters. On Fig.~\ref{fig:5}, 
\ref{fig:6}, \ref{fig:7} the one-dimensional PDF are demonstrated 
for $\Omega_{\text{int},0}$, $n$, $\Omega_{\Lambda,0}$. 

\section{Discussion of results}

We use the maximum likelihood method to estimate two basic parameters of the 
cosmology with interaction. The priors we assume for numerical computations
are: $-10<n<10$, $-1<\Omega_{\text{int}}<1$, $0<\Omega_{\text{m},0}<1$ (or 
fixed at $0.3$), and the other density parameter determined from 
$\sum_{i}\Omega_{i}=1$. From this statistical analysis we obtained that 
with assumed $\Omega_{\text{m},0}=0.3$ the interaction between ``decaying'' 
$\Lambda$ and matter sectors is negligible while the interaction between 
phantom and matter sector is required (see Table~\ref{tab:1}). The 
analogous analysis without any priors contains (see Table~\ref{tab:2}). 

To select between these models we use the Akaike and Bayesian information 
criteria (AIC and BIC) \cite{Liddle:2004nh,Szydlowski:2005qb,
Godlowski:2005tw}. These criteria were proposed to answer the question 
whether or not new parameters are relevant in the estimation. In the 
generic case introducing new parameters allows to improve the fit to 
the dataset, however information criteria try to measure whether the 
improvement in fit is large enough to consider it statistically significant. 

When we consider the interacting cosmological model we add two new parameters 
describing the interaction to the base $\Lambda$CDM model. Using the AIC 
and BIC we compare the fits of these two models to the most recent sample 
of SNIa (see Table~\ref{tab:3}). 

To improve the statistical analysis and obtain more precise 
physical results it is necessary to extend the sample to higher redshifts
$z > 1.5$. Given two mechanisms of interaction between dark 
energy and dark matter sectors, we found that the $\Lambda$CDM model is 
favored over the ``decaying'' $\Lambda$ interacting model ($\Lambda$INT) but 
it is disfavored over phantom interacting model (PhINT) for both criteria. 
The same result is obtained independently of using the prior on 
$\Omega_{\text{m},0}$. 

It is an interesting question whether it is possible to determine the direction
of energy transfer from SNIa data. To this aim we calculate the probability
that $\gamma_{0}$ is greater than zero $P(\gamma_{0}>0)$ as well as find the 
most probable values of the parameters. We have
\begin{equation}
\begin{aligned}
P(\gamma_0>0) &= 35\%    & \quad   &\text{for $\Lambda$Int models}\\
P(\gamma_0>0) &= 82.81\% & \quad   &\text{for PhInt models},
\end{aligned}
\end{equation}
under the constraint $\Omega_{\text{m},0}=0.3$. Therefore, with the confidence 
level of 95\% we can't exclude any direction of energy transfer. Because this 
result is inconclusive, we check the sign of $\gamma_{0}$ for the most 
probable values of the parameters. In both cases, we obtain (independently of 
assuming the fixed prior for $\Omega_{\text{m},0}$) that $\gamma_{0}<0$. Let 
us also note that in the case of the PhIntCDM model the values 
$\Omega_{\text{int},0}=0$ and $n=0$ are excluded with the confidence level of 
95\%.

Both the maximum likelihood method and the best fit method are used 
principally to estimate the values of parameters. Additionally the best fit 
method allows to choose the best model when a few models are to consider. 
However there is a serious shortcoming of the best fit method because models 
with greater number of parameters are generally preferred. To overcome 
this drawback and push the statistical analysis one step further we employ 
the model selection criteria. It can be seen as the second stage of 
statistical analysis after the estimation. Both of the used information 
criteria, the AIC and BIC, put some penalty on adding additional parameters 
assigning a threshold which should be exceeded to consider a model with 
a greater number of parameters to be a better fit.

\begin{table}
\caption{Results of the statistical analysis of the model obtained both for
Riess et al. samples from the best fit with minimum $\chi^2$ (denoted
with BF) and from the likelihood method (denoted with L). The same analysis
was repeated with fixed $\Omega_{\text{m},0}=0.3$ (denoted F).}
\label{tab:1}
\begin{tabular}{@{}p{1.5cm}rrrrrrr}
\hline \hline
sample & $\Omega_{\text{m},0}$ & $\Omega_{\text{int},0}$&  $n=-m-2$&
$\Omega_{\Lambda,0}$ &  $\mathcal{M}$ & $\chi^2$& method \\
\hline
$\Lambda$CDM  &  0.31 & ---  &  ---  & 0.69 &15.955&175.9 &  BF  \\
              &  0.44 & ---  &  ---  & 0.69 &15.955& ---  &  L   \\
              &F 0.30 & ---  &  ---  &F0.70 &15.955&175.9 &  BF  \\
              &F 0.30 & ---  &  ---  &F0.70 &15.955& ---  &  L   \\
\hline
$\Lambda$INT  &  0.43 & 0.28 &-10.00 & 0.29 &15.905&172.1 &  BF  \\
              &  0.44 & 0.34 & -2.70 & 0.20 &15.935& ---  &  L   \\
              &F 0.30 & 0.07 &-10.00 & 0.63 &15.925&175.2 &  BF  \\
              &F 0.30 & 0.00 &  0.00 & 0.70 &15.945& ---  &  L   \\
\hline
PhINT         &  0.46 & 0.25 &-10.00 & 0.29 &15.905&172.2 &  BF  \\
              &  0.44 & 0.26 & -2.60 & 0.62 &15.935& ---  &  L   \\
              &F 0.30 & 0.03 &  5.10 & 0.67 &15.935&173.5 &  BF  \\
              &F 0.30 & 0.03 &  0.00 & 0.67 &15.945& ---  &  L   \\
\hline
\end{tabular}
\end{table}

\begin{table}
\caption{Model parameter values obtained from the minimization procedure
carried out with the Riess et al. sample.}
\label{tab:2}
\begin{tabular}{@{}p{1.5cm}cccc}
\hline \hline
sample & $\Omega_{\text{m},0}$& $\Omega_{\text{int},0}$ &
 $\Omega_{\Lambda,0}$ & $n=-m-2$ \\
\hline
$\Lambda$CDM  & $0.31^{+0.04}_{-0.04}$ & ---
              & $0.69^{+0.04}_{-0.04}$ & --- \\
              & $0.30$ & ---
              & $0.70$ & --- \\
\hline
$\Lambda$INT  & $0.44^{+0.11}_{-0.11}$ & $0.34^{+0.39}_{-0.33}$
              & $0.20^{+0.55}_{-0.47}$ & $-2.70^{+1.80}_{-5.70}$ \\
              & $0.30$ & $0.00^{+0.26}_{-0.17}$
              & $0.70^{+0.17}_{-0.26}$ & $ 0.00^{+1.50}_{-4.80}$ \\
\hline
PhINT        & $0.44^{+0.10}_{-0.09}$ & $0.26^{+0.50}_{-0.46}$
              & $0.62^{+0.24}_{-0.76}$ & $-2.60^{+5.00}_{-3.10}$ \\
              & $0.30$ & $0.03^{+0.43}_{-0.03}$
              & $0.67^{+0.03}_{-0.43}$ & $ 0.00^{+1.50}_{-1.50}$ \\
\hline
\end{tabular}
\end{table}

\begin{table}
\caption{The values of AIC and BIC for distinguished models both for 
$\Omega_{\text{m},0}$ fitted and for $\Omega_{\text{m},0}=0.3$.}
\label{tab:3}
\begin{tabular}{c|cccc}
\hline \hline
case & AIC (fitted $\Omega_{\text{m},0}$) & AIC ($\Omega_{\text{m},0}=0.3$) 
& BIC (fitted $\Omega_{\text{m},0}$) & BIC ($\Omega_{\text{m},0}=0.3$) \\
\hline
$\Lambda$CDM   & 179.9  & 177.9 & 186.0  & 181.0  \\
$\Lambda$INT   & 180.1  & 181.2 & 192.3  & 190.4  \\
PhINT          & 180.2  & 179.5 & 192.4  & 188.7  \\
\hline
\end{tabular}
\end{table}

\acknowledgments{The paper was supported by KBN grant no. 1 P03D 003 26. 
The authors are very grateful for dr A. Krawiec and dr W. Godlowski for 
comments and discussions during the seminar on observational cosmology.}

\end{document}